\documentclass[conference]{IEEEtran}
\IEEEoverridecommandlockouts
\usepackage{cite}
\usepackage{amsmath,amssymb,amsfonts}
\usepackage{algorithmic}
\usepackage{graphicx}
\usepackage{textcomp}
\usepackage{xcolor}
\usepackage{subfig}
\usepackage{float}
\def\BibTeX{{\rm B\kern-.05em{\sc i\kern-.025em b}\kern-.08em
    T\kern-.1667em\lower.7ex\hbox{E}\kern-.125emX}}

\begin{document}

\title{Implementation and Evaluation of Physical Layer Key Generation on SDR based LoRa Platform
}
\author{
	\IEEEauthorblockN{
		Yingying Hu,
		Dongyang Xu$^{
			\ast}$, and
		Tiantian Zhang}
	\IEEEauthorblockA{School of Information and Communications Engineering, Xi'an Jiaotong University, Xi'an, 710049, China}
	
	\IEEEauthorblockA{Email: \textit{2176801644@stu.xjtu.edu.cn, xudongyang@xjtu.edu.cn, tiantianzhang@stu.xjtu.edu.cn}}
}

\maketitle

\begin{abstract}
Physical layer key generation technology which leverages channel randomness to generate secret keys has attracted extensive attentions in long range (LoRa)-based networks recently. We in this paper develop a software-defined radio (SDR) based LoRa communications platform using GNU Radio on universal software radio peripheral (USRP) to implement and evaluate typical physical layer key generation schemes. Thanks to the flexibility and configurability of GNU Radio to extract LoRa packets, we are able to obtain the fine-grained channel frequency response (CFR) through LoRa preamble based channel estimation for key generation. Besides,  we propose a low-complexity preprocessing method to enhance the randomness of quantization while reducing the secret key disagreement ratio. The results indicate that we can achieve 367 key bits with a high level of randomness through just a single effective channel probing in an indoor environment at a distance of 2 meters under the circumstance of a spreading factor (SF) of 7, a preamble length of 8, a signal bandwidth of 250 kHz, and a sampling rate of 1 MHz.
\end{abstract}


\section{Introduction}
LP-WAN (Low-Power Wide-Area Network) has attracted lots of attentions in future communications due to its massive machine connectivity, extended coverage, low-power consumption and many others. LoRa (Long Range Radio)  communications, one of the prominent LP-WAN communications technologies, is also additionally featured by open standard and adaptive data rates in unlicensed frequency bands, making it well-suited for various applications in smart cities. Numerous LoRa nodes, including IoT devices (EDs) and other sensors such as electric meters, smoke alarms or temperature and humidity sensors, can communicate with a far away gateway which can be accessed by up to tens of thousands of LoRa end nodes so that meeting the demands for extensive IoT connectivity.\\
\indent Meawhile, with the continuous increase in connected devices, ensuring communication security has become an issue that cannot be ignored. However, due to the inherent broadcasting nature of wireless communication and long-distance transmission, LoRa packets have longer duration in the air, making it more susceptible to various attacks such as eavesdropping and interference\cite{b1}. The traditional encryption method used in LoRa relies on advanced encryption standard (AES) algorithm in upper layer. Unfortunately, due to limited computational capability and the difficulty of large-scale key generation and management, it is not suitable for a LoRa-based network when in practical applications. An efficient and lightweight security encryption method is required to ensure the transmission security of LoRa. \\
\indent The physical layer security (PLS) scheme utilizes the reciprocity, randomness and decorrelation of the wireless channel to extract channel state information (CSI), channel frequency response (CFR) and received signal strength indicator (RSSI) for key generation. Due to its advantages of low complexity and lightweight nature, it has sparked extensive research in various scenarios\cite{b2}--\cite{b5}. Currently, there have been some studies conducted on the PLS scheme in LoRa-based networks. Zhang \textit{et al}.\cite{b6} employed a differential quantization method to extract high levels of randomness from RSSI for LoRa key generation. Xu \textit{et al}.\cite{b7} first proposed a comprehensive secret key generation protocol for LoRa-based networks using RSSI, namely LoRa-Key. This work employs a compressive sensing-based reconciliation approach to address the mismatched bits generated between the gateway and LoRa nodes. Junejo \textit{et al}.\cite{b8} proposed the LoRa-LiSK scheme and achieved the first implementation of key generation in static indoor-to-outdoor scenario. LoRa-LiSK employs some preprocessing techniques, including the Savitzky-Golay filter, to mitigate the issue of correlation reduction caused by long-distance transmission. It also adopts multi-level quantization to enhance the key generation rate. Zhang \textit{et al}.\cite{b9} first defined the LoRa constellation mapping and introduced a novel physical layer encryption algorithm under RSSI-based LoRa key generation. This algorithm doesn't simply perform a XOR operation between the generated key and the LoRa frame. Instead, it randomly selects a continuous sequence of key bits to encrypt each symbol within the LoRa frame.\\ 
\indent However, these studies are all based on LoRa hardware devices, and due to the current LoRa driver specifications, LoRa hardware devices can only offer coarse-grained channel information like RSSI. Consequently, a single channel probing can only yield one piece of data, necessitating multiple probes to collect sufficient data, which in turn, leads to inevitable communication delays and limited secret key generation rate. However, with the development of software-defined radio (SDR), a significant portion of hardware circuit functions can be transitioned to software implementation. This offers a solution for obtaining fine-grained CFR values in LoRa transmissions.\\
\indent This paper aims to study methods for secret key generation utilizing CFR in LoRa-based networks. We leverage the flexibility and configurability of SDR to implement a LoRa network using GNU Radio and universal software radio peripheral (USRP), and we extract the LoRa packets received by the USRP in GNU Radio. Then, according to the frame structure characteristics of LoRa, we use a preamble-based channel estimation method to obtain CFR. we also employ a simple preprocessing method to enhance the randomness of quantization and reduce the secret key disagreement ratio. By analyzing the performance of our key generation, it demonstrates that, the secret key generation rate, the secret key disagreement ratio and the randomness of the key in our SDR-based LoRa system are greatly improved. Under specific parameter configurations, it is possible to achieve a 367-bit key through a single channel probing in a real indoor environment at a distance of 2 meters, which is much higher than current RSSI-based key generation.
\section{System Model} 
\subsection{System Overview}
In this paper, we consider a key generation model in a LoRa-based network as shown in Fig. 1, which includes a gateway and two LoRa nodes. LoRa-A serves as the legitimate user, whereas LoRa-B is an illegitimate eavesdropper. Both LoRa-A and the Gateway send normal LoRa frames to each other at the same frequency in a real wireless channel. Subsequently, channel estimation is performed individually to acquire the CFR values for the secret key generation. We assume that LoRa-B is sufficiently distant from both LoRa-A and the Gateway, and it only passively eavesdrops from the Gateway. Therefore, channel measurements from LoRa-A to the Gateway and from the Gateway to LoRa-A are unavailable for LoRa-B.
\begin{figure}[htbp]
	\centerline{\includegraphics[width=0.45\textwidth]{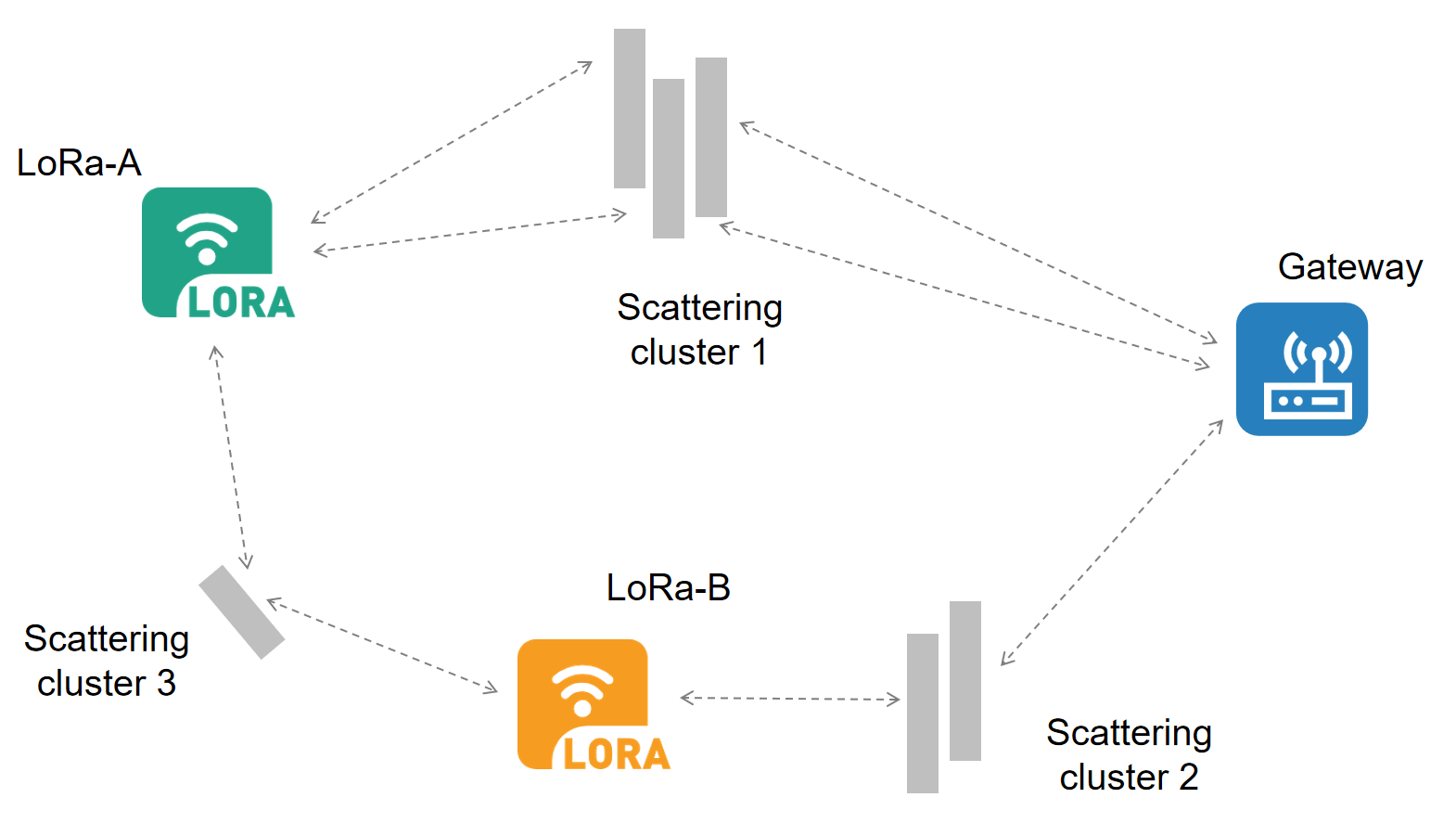}}
	\caption{Secret key generation model in a LoRa-based network.}
	\label{fig}
\end{figure}
\\ \indent Similar to all physical-layer key generation methods, the process consists of four components, namely channel probing, quantization, information reconciliation and privacy amplification\cite{b10}, as shown in Fig. 2. Channel probing is used to extract channel parameters from the wireless channel. In this study, CFR values are utilized. Quantization converts CFR values into the initial binary key bits, and in this paper, we employ an adaptive dual-threshold quantization method. Prior to quantization, we propose a simple preprocessing technique to enhance the randomness of quantization. Due to the fact that the channel not being fully reciprocal and the influence of noise, the initial key bits of the two legitimate parties may contain inconsistent bits. Information reconciliation is employed to correct these mismatched bits, and the information reconciliation protocol cascade\cite{b11} is used in this paper. And finally, we perform privacy amplification to prevent key leakage. Each phase will be introduced in detail as follow.
\begin{figure}[htbp]
	\centerline{\includegraphics[width=0.45\textwidth]{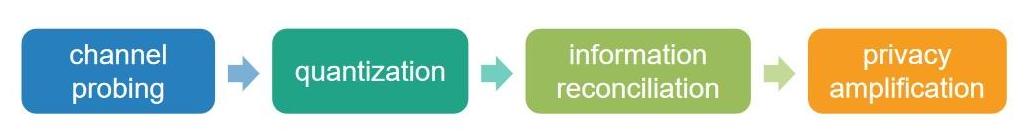}}
	\caption{Steps for key generation.}
	\label{fig}
\end{figure}
\subsection{Channel Probing}
In the channel probing phase, LoRa-A and the Gateway send LoRa frames to each other. Then, the received frames are used for channel estimation to acquire the CFR values. A typical LoRa frame consists of preamble, header, and data payload. Savaux et al.\cite{b12} have proposed a preamble-based channel estimation method for LoRa and demonstrated that channel estimation methods designed for cyclic prefix-orthogonal frequency division multiplexing (CP-OFDM), such as least squares (LS) and minimum mean square error (MMSE), can be adapted for LoRa channel estimation. In this paper, we choose the most commonly used LS estimator.\\
\indent Due to the flexibility of GNU Radio, we can extract the preamble from the received LoRa frames. The preamble of a LoRa frame consists of $K$ symbols, where $K$ is usually set to 8. Each symbol is an upchirp, which can be expressed in the time domain as:
\begin{equation}
	s(t)=\exp (j \pi(-B W+k t) \cdot t) \label{eq:einstein}
\end{equation}
\indent Where $BW$ is the bandwidth of the LoRa signal, $k=\frac{B W}{T}$ represents the sweep rate, $t \in[0, T]$ and $T$ is the duration of an upchirp. In the following, we will focus on CFR estimation on the Gateway’s side.\\ 
\indent LoRa-A sends a LoRa frame to the Gateway, one of the preamble symbols $\mathrm{r}_{\mathrm{G}}(\mathrm{t})$, in the received signal at the Gateway can be obtained in the time domain:
\begin{equation}
	\mathrm{r}_{\mathrm{G}}(\mathrm{t})=\mathrm{h}_{\mathrm{AG}} \mathrm{s}_{\mathrm{G}}[\mathrm{t}]+\mathrm{n}_{\mathrm{G}}[\mathrm{t}] \label{eq:einstein}
\end{equation}
\indent Where $\mathrm{s}_{\mathrm{G}}[\mathrm{t}]$ is one preamble symbol of the transmitted LoRa frame, $\mathrm{h}_{\mathrm{AG}}$ is the channel matrix from LoRa-A to the Gateway, and $\mathrm{n}_{\mathrm{G}}[\mathrm{t}]$ corresponds to the noise at the Gateway. Through Fourier transformation, we can obtain:
\begin{equation}
	\mathrm{R}_{\mathrm{G}}(\mathrm{t})=\mathrm{H}_{\mathrm{AG}} \mathrm{S}_{\mathrm{G}}[\mathrm{t}]+\mathrm{N}_{\mathrm{G}}[\mathrm{t}] \label{eq:einstein}
\end{equation}
\indent Where $\mathrm{R}_{\mathrm{G}}(\mathrm{t})$, $\mathrm{H}_{\mathrm{AG}}$, $\mathrm{S}_{\mathrm{G}}[\mathrm{t}]$ and $\mathrm{N}_{\mathrm{G}}[\mathrm{t}]$ represent the frequency domain representations for $\mathrm{r}_{\mathrm{G}}(\mathrm{t})$, $\mathrm{h}_{\mathrm{AG}}$, $\mathrm{s}_{\mathrm{G}}[\mathrm{t}]$ and $\mathrm{n}_{\mathrm{G}}[\mathrm{t}]$, respectively. We choose the LS estimator to estimate the CFR values:
\begin{equation}
	\widehat{\mathrm{H}}_{\mathrm{AG}}=\mathrm{R}_{\mathrm{G}}(\mathrm{t}) \mathrm{S}_{\mathrm{G}}[\mathrm{t}]^{-1}=\mathrm{H}_{\mathrm{AG}}+\mathrm{N}_{\mathrm{G}}[\mathrm{t}] \mathrm{S}_{\mathrm{G}}[\mathrm{t}]^{-1} \label{eq:einstein}
\end{equation}

\indent Where $\mathrm{S}_{\mathrm{G}}[\mathrm{t}]^{-1}$ is the  inverse matrix of $\mathrm{S}_{\mathrm{G}}[\mathrm{t}]$. We know that although the LS estimate is simple, it tends to amplify noise. In order to improve the LS estimate, we make full use of $K$ consecutive preamble symbols to average the noise. Consequently, we obtain the CFR values for the Gateway as:
\begin{equation}
	\overline{\mathrm{H}}_{\mathrm{AG}}=\frac{1}{\mathrm{~K}} \sum_{\mathrm{i}=1}^{\mathrm{K}} \mathrm{H}_{\mathrm{AG}}^{(\mathrm{i})} \label{eq:einstein}
\end{equation}
\indent Where $\mathrm{H}_{\mathrm{AG}}^{(\mathrm{i})}$ is the CFR estimated using the i-th preamble symbol in the LoRa frame. LoRa-A employs the same method to estimate its CFR values as:
\begin{equation}
	\overline{\mathrm{H}}_{\mathrm{GA}}=\frac{1}{\mathrm{~K}} \sum_{\mathrm{i}=1}^{\mathrm{K}} \mathrm{H}_{\mathrm{GA}}^{(\mathrm{i})} \label{eq:einstein}
\end{equation}
\indent $\overline{\mathrm{H}}_{\mathrm{AG}}$ and $\overline{\mathrm{H}}_{\mathrm{GA}}$ exhibit a high degree of correlation and can be used for subsequent key generation steps.
\subsection{Adaptive Dual-threshold Quantization}
In this paper, we quantize the amplitude of the CFR into binary bits. Next, we will introduce the adaptive dual-threshold quantization algorithm in detail:
\begin{enumerate}
\item [1.] LoRa-A and the Gateway independently partition their CFR amplitude values $\mathrm{H}_{\mathrm{A}}$ and $\mathrm{H}_{\mathrm{G}}$ into blocks of length $m$ and compute the thresholds $q_i^{+}$ and $q_i^{-}$ for each block:
\begin{equation}
	\begin{aligned}
		& q_i^{+}=\operatorname{mean}\left(H^i\right)+\alpha \cdot \sigma\left(H^i\right) \\
		& q_i^{-}=\operatorname{mean}\left(H^i\right)-\alpha \cdot \sigma\left(H^i\right)
	\end{aligned} \label{eq:einstein}
\end{equation}
Where $H^i$ is the i-th block of the amplitude values for either LoRa-A or the Gateway, $\operatorname{mean}\left(\right)$ represents mean calculation, $\sigma\left(\right)$ is variance calculation, and $\alpha$ is a threshold adjustment factor.
\item [2.] LoRa-A searches for amplitude values in each block that are smaller than $q_i^{+}$ and larger than $q_i^{-}$, then sends a list of corresponding indices for these values to the Gateway.
\item [3.] The Gateway removes amplitude values smaller than $q_i^{+}$ and larger than $q_i^{-}$ from each of its blocks, as well as the values corresponding to the indices in the received list. Then, Gateway sends the list of corresponding indices for the removed values back to LoRa-A.
\item [4.] LoRa-A removes the amplitude values according to the list received from the Gateway.
\item [5.] LoRa-A and the Gateway quantize the rest of amplitude values into binary bits using traditional Gray code :
\begin{equation}
	\mathrm{Q}(\mathrm{x})= \begin{cases}1 & \mathrm{x}>=q_i^{+} \\ 0 & \mathrm{x}<=q_i^{-}\end{cases}
	\label{eq:einstein}
\end{equation}
\end{enumerate}
\indent\indent However, traditional Gray code will lead to the generation of too many consecutive 0 or 1, which reduces the randomness of the key. Reference [4] has proposed the double-Gray (D-Gray) code to tackle this issue, wherein 0 is replaced by 01 and 1 is replaced by 10. Although it overcomes the disadvantage of traditional Gray code, it comes at the cost of increased complexity in information reconciliation. Because one mismatched bit in the traditional Gray code will transforms into two mismatched bits in the D-Gray code. Consequently, additional rounds of information reconciliation are required to correct the extra mismatched bits.\\
\indent We propose to shuffle the CFR values using the same rule at both two parties before quantization, which makes a more uniformly distributed of the CFR amplitude values, resulting in an increase of randomness, thereby preventing the occurrence of a lot of continuous 0 or 1. 
\begin{figure}[H]
	\centerline{\includegraphics[width=0.45\textwidth]{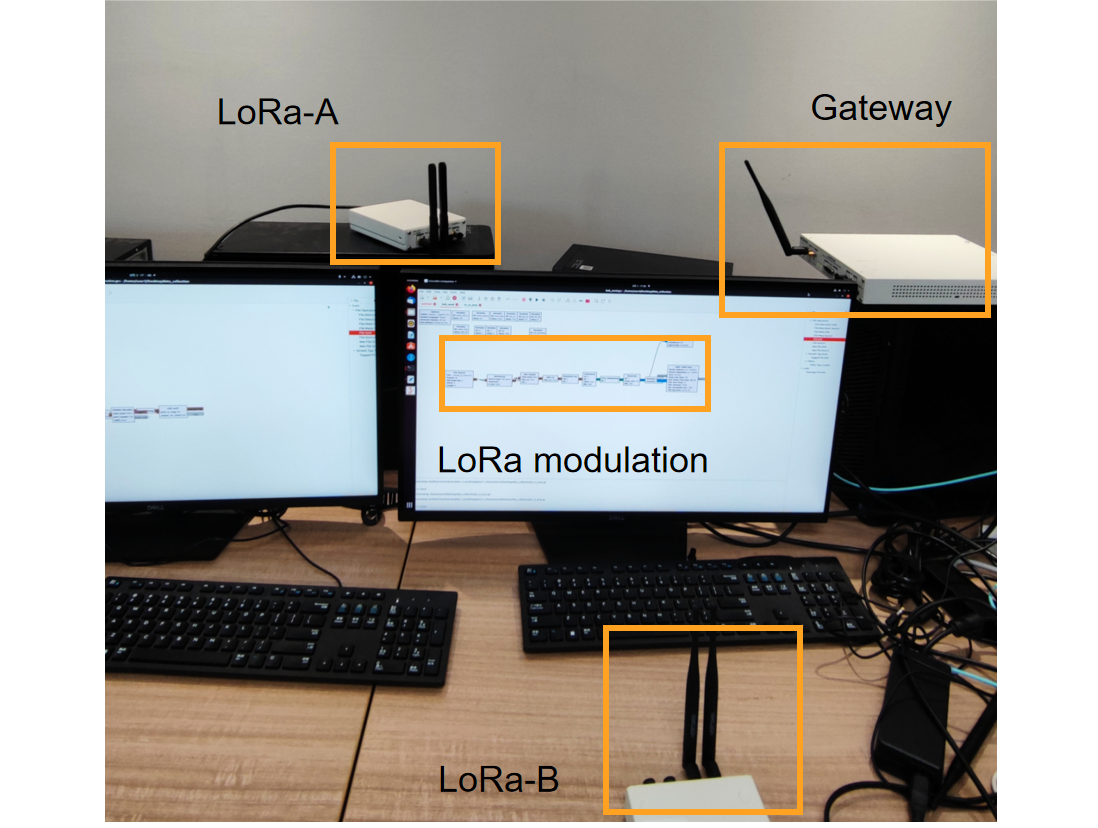}}
	\caption{LoRa-based system.}
	\label{fig}
\end{figure}
\subsection{Information Reconciliation}
After quantization, both LoRa-A and the Gateway generate an initial key, which might not be entirely identical. In this paper, the information 
reconciliation protocol cascade is employed to address this promble. Due to it is not the focus of this paper and space limitations, readers can refer to the reference \cite{b11} for a further understanding of this protocol.
\subsection{Privacy Amplification}
In the privacy amplification phase, LoRa-A and the Gateway use the key obtained after information reconciliation to run an SHA-256 process separately, which will return a digest. Subsequently, the digests are exchanged between LoRa-A and the Gateway. If their digests are the same, the key will be tested by NIST test. If passing the NIST test, the key can be used for the encryption process. 
\section{Results and Performance Evaluation}
\subsection{Experimental Setup}
As shown in Fig. 3, the entire system consists of three computers connected to the USRP through the USRP hardware driver (UHD), each equipped with the LoRa physical layer implementation under GNU Radio within the Ubuntu environment. These three computers serve as legitimate user LoRa-A, eavesdropper LoRa-B, and the Gateway, respectively. The computer is mainly responsible for the encoding and decoding of LoRa and the modulation and demodulation. For the configuration of LoRa modulation parameters, we set the preamble length to 8, the spreading factor (SF) to 7, the signal bandwidth to 250 kHz, the center frequency to 868 MHz, and the sampling rate to 1 MHz. We transmit arbitrary information that is modulated by LoRa in GNU Radio and then sent via the USRP at the sending end. At the receiving end, we use the File Sink module provided by GNU Radio to extract the preamble from the LoRa frame received by USRP, which is subsequently employed for the estimation of CFR values.
\begin{figure}[H]
	\centerline{\includegraphics[width=0.45\textwidth]{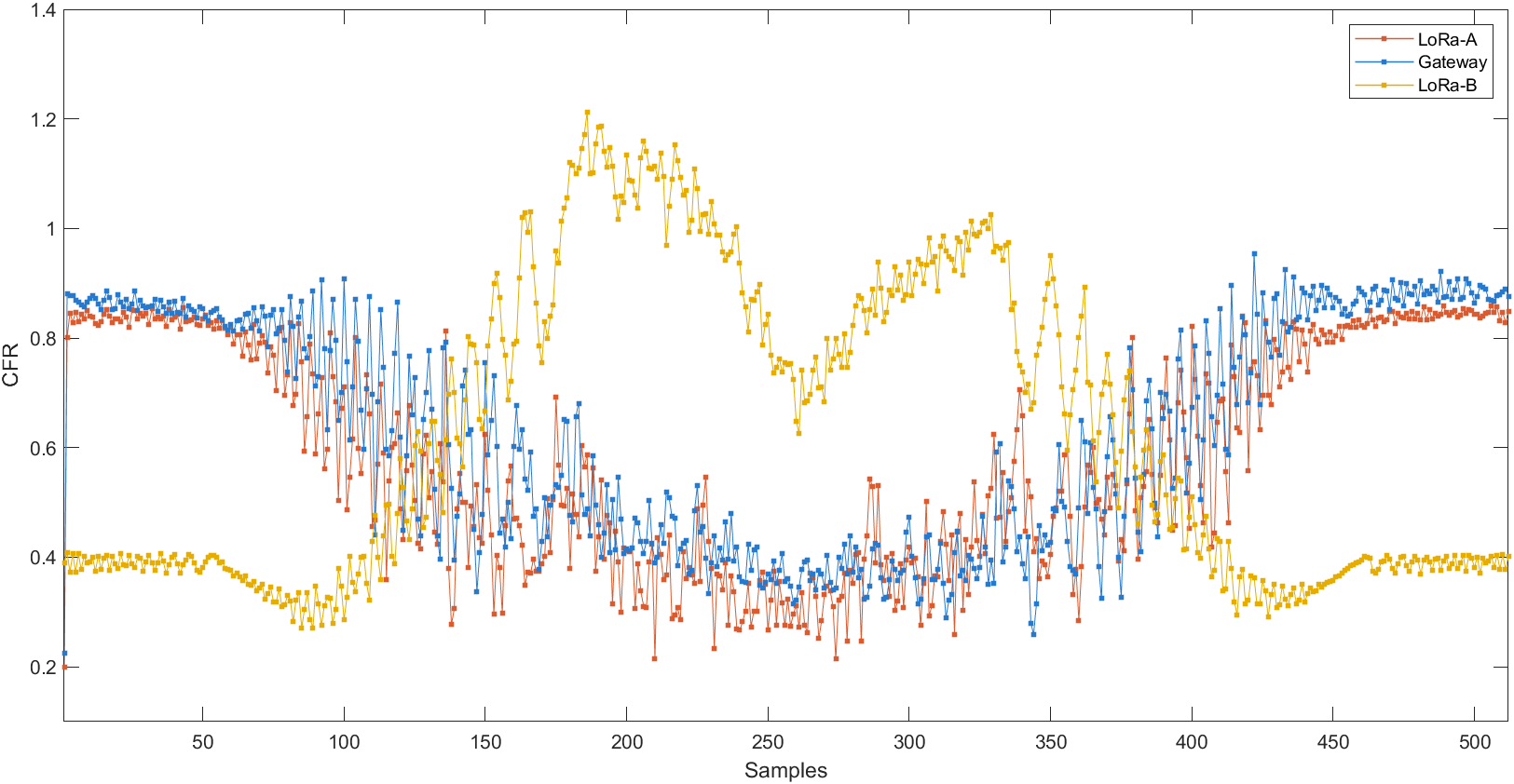}}
	\caption{CFR values of LoRa-A, LoRa-B and the Gateway in an indoor environment.}
	\label{fig}
\end{figure}
\begin{figure}[H]
		\centering 
		\vspace{-0.7cm}  
		\subfloat[]   
		{
			\label{fig:subfig1}\includegraphics[width=0.225\textwidth]{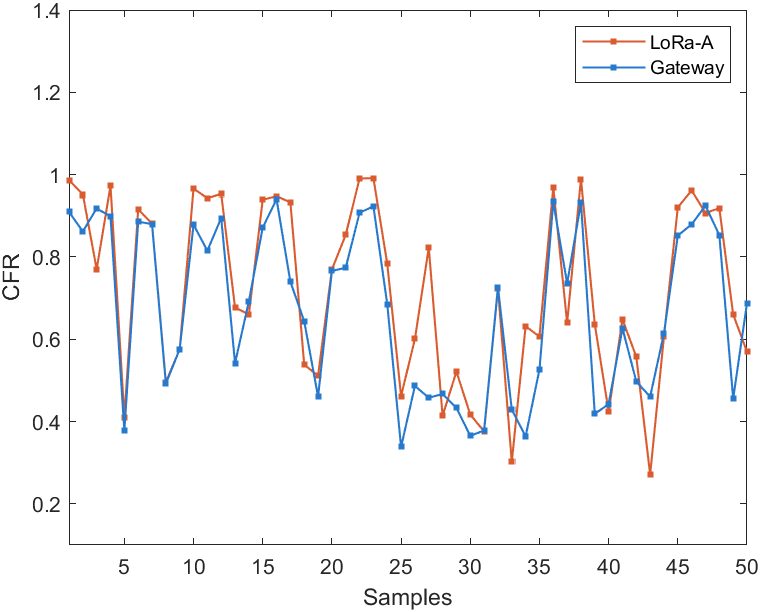}
		}
		\subfloat[]
		{
			\label{fig:subfig2}\includegraphics[width=0.225\textwidth]{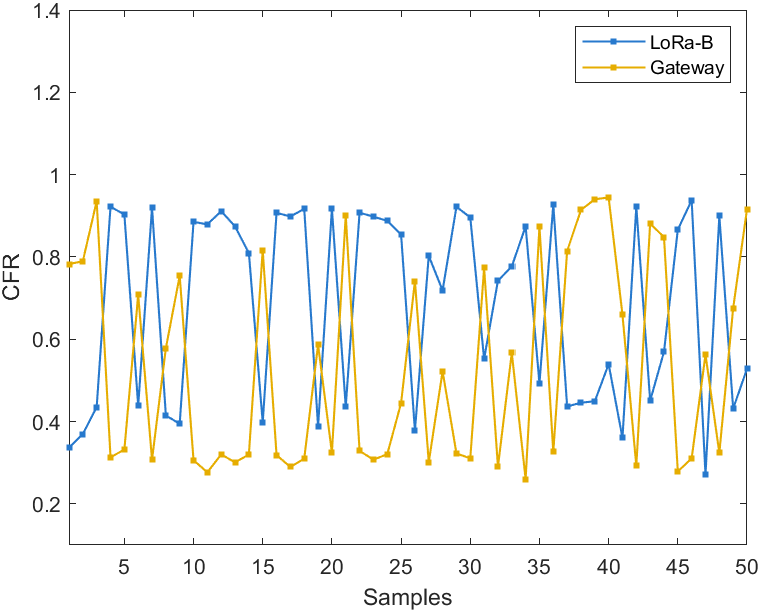}
		}
		\caption{Shuffled CFR values of LoRa-A, LoRa-B, and the Gateway. (a) CFR values between LoRa-A and the Gateway. (b) CFR values between LoRa-B and the Gateway.}    
		\label{fig:subfig_1}           
\end{figure}
\indent We place these three computers in an indoor environment with stationary obstacles at a distance of two meters. The amplitudes of CFR values obtained by LoRa-A, LoRa-B and the Gateway through channel probing are depicted in Fig. 4. It can be seen that the amplitudes of CFR values obtained by LoRa-A and the Gateway exhibit a high degree of similarity, while the CFR amplitudes of LoRa-B show significant differences from both LoRa-A and the Gateway. After being randomly shuffled, the partial CFR amplitudes of both LoRa-A and the Gateway are shown in Fig. 5 (a). It is evident that random shuffling substantially enhances randomness while maintaining a high degree of correlation. We assume that LoRa-B knows the rule of the random shuffling and applies it to its own CFR amplitudes. As illustrated in Fig. 5 (b), although the CFR amplitudes of LoRa-B also demonstrate increased randomness, they still retain noticeable distinctions from those of the Gateway.
\subsection{Evaluation Metrics}
\begin{itemize}
\item SKDR (secret key disagreement ratio): Although LoRa-A and the Gateway observe the same random source, differences in background noise and hardware will lead to disparities between the CFR values estimated by both sides. Additionally, due to the presence of a certain time interval in channel probing between LoRa-A and the Gateway, especially when the LoRa signal transmits over long distances, the probing interval significantly increases, resulting in a deterioration of channel reciprocity, which leads to inconsistencies in the quantized key bits. Therefore, SKDR is used to assess the inconsistency rate of the initial key bits obtained by both sides. SKDR is defined as follows:
\begin{equation}
	\mathrm{\mathrm{Q}}=\frac{\sum key_{A} \oplus key_{G}}{N} 
	\label{eq:einstein}
\end{equation} 
Where $key_{A}$ and $key_{G}$ represent the initial key bits obtained after quantization by LoRa-A and the Gateway, respectively. $N$ is the length of the initial key bits, and $\oplus$ denotes the XOR operation.
\item SKGR (secret key generation rate): SKGR describes the rate at which both parties generate the key bits. It is defined as the amount of key bits generated per second or the amount of key bits generated per channel probing. In this paper, it is defined as the latter.
\item Randomness: The generated secret key bits require a high level of randomness, making them  difficult to be predicted by potential attackers. We employ several NIST\cite{b13} tests to assess the randomness of the secret key bits generated by our method.
\end{itemize}
\subsection{Performance Evaluation}
In this section, we begin by comparing the effects of the threshold adjustment factor $ \alpha$ and the block size $m$ in the dual-threshold quantization process on both original key generation method and our method with random shuffling preprocessing. Subsequently, we proceed to evaluate the performance of our method.
\begin{figure}[H]
	\centerline{\includegraphics[width=0.4\textwidth]{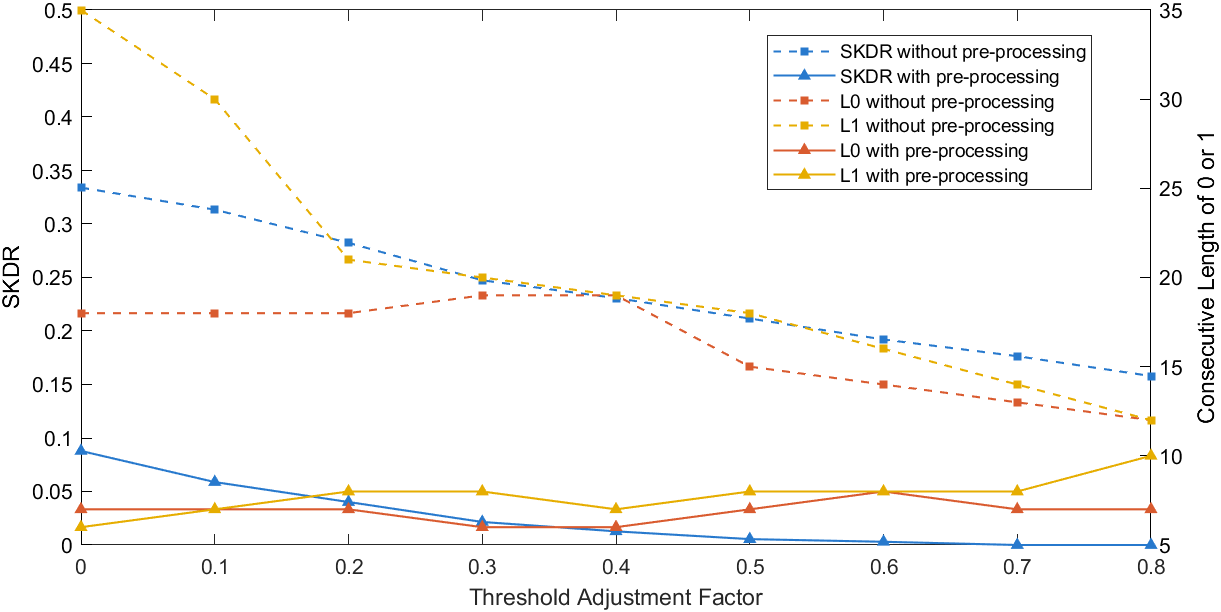}}
	\caption{Impact of threshold adjustment factor $\alpha$.}
	\label{fig}
\end{figure} 
\paragraph{Influence of Threshold Adjustment Factor}With the block size $m$ fixed at 64, as depicted in Fig. 6 where $L_0$ and $L_1$ respectively indicate the maximum consecutive length of 0 and 1, it is evident that $L_0$, $L_1$ and SKDR demonstrate a decreasing trend with the increase of the threshold adjustment factor $\alpha$ when without preprocessing, while a very small change of them when use preprocessing. Our optimized method exhibits superior adaptability to changes in the threshold adjustment factor.
\paragraph{Influence of Block Size}With the the threshold adjustment factor fixed at 0.5, as illustrated in Fig. 7, the results show that an increase in the block size $m$ results in an increase in $L_0$ and $L_1$  while a decrease in SKDR when using original method. Conversely, minimal fluctuations are observed in them when preprocessing is employed. The optimized method we developed demonstrates enhanced adaptability to variations in quantization block size.
\begin{figure}[H]
	\centering 
	{\includegraphics[width=0.4\textwidth]{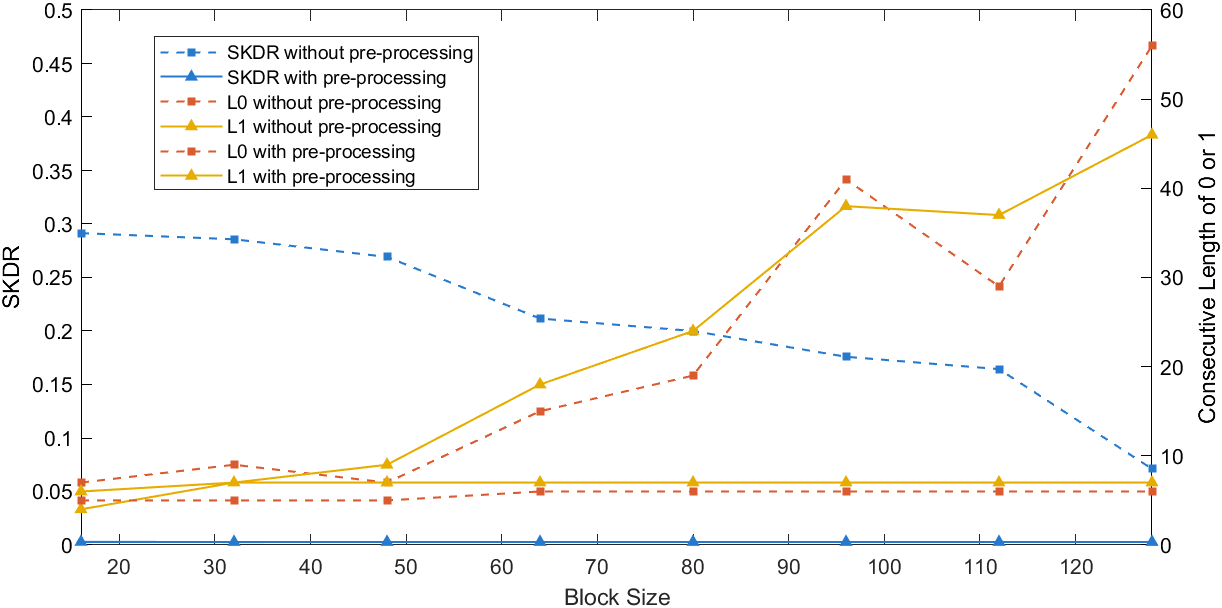}}
	\caption{Impact of block size $m$.}
	\label{fig}
\end{figure}
\paragraph{Secret Key Disagreement Ratio}From Fig. 6 and Fig. 7, we can observe that our method consistently yields a significantly lower secret key disagreement ratio compared to the original method, across various quantization block sizes and threshold adjustment factors. For instance, when the threshold adjustment factor is 0.1 and the block size is 64, the SKDR in the original method exceeds 0.35, while it remains below 0.1 in our method. Furthermore, with a threshold adjustment factor of 0.5 and any block size, the SKDR in our method nearly approaches 0. Overall, our solution effectively improves the SKDR. 
\paragraph{Secret Key Generation Rate}With a threshold adjustment factor of 0.5 and a block size of 64, our CFR-based key generation method achieves up to 367 key bits in a single effective channel probing, while maintaining a SKDR below 0.05. In contrast, only 189 key bits can be obtained in original method, with a SKDR greater than 0.21. Our solution obtains a higher quantity of key bits at a low SKDR, which reduces the latency caused by multiple channel probing and the complexity of subsequent information reconciliation.
\begin{table}[htbp]
	\caption{P-values of NIST Statistical Test}
	\begin{center}
		\begin{tabular}{|l|c|}
			\hline
			NIST Test&p-value\\
			\hline
			Frequency&0.376759\\
			FFT Test&0.144292\\
			Longest Run&0.285956\\
			Linear Complexity&0.808840\\
			Block Frequency&0.376759\\
			Cumulative Sums&0.737518\\
			Approximate Entropy&1.000000\\
			Non Overlapping Template&0.116417\\
			\hline
		\end{tabular}
		\label{tab1}
	\end{center}
\end{table}
\paragraph{Randomness}To validate the randomness of the final key, we assess its randomness using the NIST\cite{b13} suite of statistical tests. The P-values corresponding to different tests are presented in Table 1. Traditionally, if the p-value is less than 0.01, the random hypothesis is rejected, suggesting that the key is not random. We can observe that all the P-values are greater than 0.01 in sense that our key possesses a high degree of randomness.
\section{CONCLUSION}
In this paper, we build a SDR based LoRa network using GNU Radio and USRP, and realize CFR-based LoRa key generation and performance evaluation on this platform. Under the configured LoRa parameters, we achieve 367 key bits with just one effective channel probing in an indoor environment at a distance of 2 meters, substantially boosting the key generation rate compared to LoRa hardware-based key generation. Our work demonstrates that SDR offers new approach for the currently constrained LoRa key generation. For instance, we can leverage SDR to extract data packets from any node in the LoRa physical layer, enabling the study of algorithms for obtaining more channel randomness. We can also conveniently modify LoRa physical layer parameters through SDR to evaluate key generation performance under different LoRa configurations. This holds significance as a reference for future hardware-based LoRa key generation. In the future, our work will further explore LoRa key generation in scenario of long-distance transmission and complex environments based on this work.

\end{document}